# Magnetic properties of spin-3/2 Blume–Capel model on the hexagonal Ising nanowire

Yusuf Kocakaplan[1] and Mehmet Ertaş[2,*]

[1]*Graduate School of Natural and Applied Sciences, Erciyes University, 38039 Kayseri, Turkey*
[2]*Department of Physics, Erciyes University, 38039 Kayseri, Turkey*

## Abstract

Magnetic properties, such as magnetizations, internal energy, specific heat, entropy, Helmholtz free energy and phase diagrams of the spin-3/2 Blume-Capel model on hexagonal Ising nanowire (HIN) with core-shell structure are studied by utilizing the effective-field theory (EFT) with correlations. Moreover, the hysteresis behaviors of the system are also investigated and the effects of Hamiltonian parameters on hysteresis behaviors are discussed, in detail. Finally, the obtained results are compared with some experimental and theoretical results and a qualitatively good agreement is found.

## 1. Introduction

In recent years, there has been a growing attention in magnetic behaviors of nanostructured materials [1, 2]. The reason of this fact is that they have unusual properties compared with bulk materials and have an important potential technological applications in various areas, such as they can be used for nonlinear optics [3], sensors [4], environmental remediation [5], permanent magnets [6], magnetic recording media [7], and also used for bio-separation [8, 9]. These facts give a strong motivation for the investigation of magnetic nanoparticles systems such as nanowires and nanotubes. Theoretically, in statistical mechanics and condensed matter physics magnetic properties of these materials are actively studied by using different methods, such as mean-field approximation (MFA) [10], effective-field theory (EFT) [11-17], Monte-Carlo simulations (MCs) [18-21], Bethe-Pierls approximation [22] and Green's function technique [23, 24]. On the other hand, recently, some interest has been directed towards on understanding of more complicated nanostructured materials with a spin higher than one and mixed spins. Some characteristic behaviors of magnetic Ising nanotubes have been studied for the spin-3/2 [25], mixed spins (1/2-1) [26, 27] and mixed spins (3/2-1) [28] Ising models. Moreover, some magnetic properties have been investigated for mixed spins (1/2-1) [29, 30], mixed spins (3/2-1) [31] HIN system and ternary Ising spins (1/2, 1, 3/2) [32] for nanoparticle. We should also mention that dynamic behaviors of nanostructured materials have been studied by using EFT [33-36] and MFT based on Glauber dynamics [37, 38] and MCs [39, 40].

---
[*] Corresponding author. Tel. +90 352 2076666; Fax: +90 352 4374901.
E-mail addresses: mehmetertas@erciyes.edu.tr (Mehmet Ertaş)

Despite these studies, as far as we know, the magnetic properties and phase diagrams of a spin-3/2 HIN system with core-shell structure are not studied by utilizing the any method. Therefore, in this work, we investigate the magnetic properties and phase diagrams of a spin-3/2 HIN system with core-shell structure by utilizing the EFT with correlations.

The outline of the paper is as follows. In Sec. 2, the EFT formalism is presented briefly. The detailed numerical results and discussions are given in Section 3. Finally, Section 4 is devoted to a summary and conclusion.

## 2. Formalism

The Hamiltonian of the cylindrical spin-3/2 HIN system includes nearest-neighbor interactions and the crystal field is given as follows

$$H = -J_S \sum_{\langle ij \rangle} S_i S_j - J_C \sum_{\langle mn \rangle} S_m S_n - J_1 \sum_{\langle im \rangle} S_i S_m - D \left( \sum_i (S_i)^2 + \sum_m (S_m)^2 \right) - h \left( \sum_i S_i + \sum_m S_m \right), \quad (1)$$

where each $S_i$ can take the values $\pm 3/2$ and $\pm 1/2$ and $\langle \ldots \rangle$ indicates summation over all pairs of nearest-neighbor sites. The exchange interaction parameters $J_S$, $J_C$ and $J_1$ are interactions between the two nearest-neighbor magnetic atoms at the shell, core and between the shell and core, respectively (see Fig. 1). D stands for the single-ion anisotropy, i.e. the crystal field, and h is the external magnetic field. The surface exchange interaction $J_S = J_C (1 + \Delta_S)$ and interfacial coupling $r = J_1 / J_C$ are often defined to clarify the effects of the surface and interfacial exchange interactions on the physical properties in the nanosystem, respectively.

Within the EFT with correlations framework, one can easily find the magnetizations $M_S$, $M_C$, the quadruple moments $q_S$, $q_C$ and the octupolar moments $r_S$, and $r_C$ as coupled equations, for the cylindrical spin-3/2 HIN system as follows:

$$M_S = \langle S_i \rangle = \left[ a_0 + a_1 \langle S_i \rangle + a_2 \langle S_i \rangle^2 + a_3 \langle S_i \rangle^3 \right]^4 \left[ b_0 + b_1 \langle S_m \rangle + b_2 \langle S_m \rangle^2 + b_3 \langle S_m \rangle^3 \right] F_1(x+h) \Big|_{x=0}, \quad (2a)$$

$$M_C = \langle S_m \rangle = \left[ c_0 + c_1 \langle S_m \rangle + c_2 \langle S_m \rangle^2 + c_3 \langle S_m \rangle^3 \right]^2 \left[ b_0 + b_1 \langle S_i \rangle + b_2 \langle S_i \rangle^2 + b_3 \langle S_i \rangle^3 \right]^6 F_1(x+h) \Big|_{x=0}, \quad (2b)$$

$$q_S = \langle S_i \rangle^2 = \left[ a_0 + a_1 \langle S_i \rangle + a_2 \langle S_i \rangle^2 + a_3 \langle S_i \rangle^3 \right]^4 \left[ b_0 + b_1 \langle S_m \rangle + b_2 \langle S_m \rangle^2 + b_3 \langle S_m \rangle^3 \right] F_2(x+h) \Big|_{x=0}, \quad (3a)$$

$$q_C = \langle S_m \rangle^2 = \left[ c_0 + c_1 \langle S_m \rangle + c_2 \langle S_m \rangle^2 + c_3 \langle S_m \rangle^3 \right]^2 \left[ b_0 + b_1 \langle S_i \rangle + b_2 \langle S_i \rangle^2 + b_3 \langle S_i \rangle^3 \right]^6 F_2(x+h) \Big|_{x=0}, \quad (3b)$$

$$r_S = \langle S_i \rangle^3 = \left[ a_0 + a_1 \langle S_i \rangle + a_2 \langle S_i \rangle^2 + a_3 \langle S_i \rangle^3 \right]^4 \left[ b_0 + b_1 \langle S_m \rangle + b_2 \langle S_m \rangle^2 + b_3 \langle S_m \rangle^3 \right] F_3(x+h) \Big|_{x=0}, \quad (4a)$$

$$r_C = \langle S_m \rangle^3 = \left[ c_0 + c_1 \langle S_m \rangle + c_2 \langle S_m \rangle^2 + c_3 \langle S_m \rangle^3 \right]^2 \left[ b_0 + b_1 \langle S_i \rangle + b_2 \langle S_i \rangle^2 + b_3 \langle S_i \rangle^3 \right]^6 F_3(x+h) \Big|_{x=0}, \quad (4b)$$

while the $a_i, b_i$ and $c_i$ coefficients are given in the Appendix. The functions $F_1(x)$, $F_2(x)$ and $F_3(x)$ in Eqs. (2)-(4) are given as

$$F_1(x+h) = \frac{1}{2} \frac{3\sinh[3\beta(x+h)/2] + \sinh[\beta(x+h)/2]\exp(-2\beta D)}{\cosh[3\beta(x+h)/2] + \cosh[\beta(x+h)/2]\exp(-2\beta D)}, \tag{5a}$$

$$F_2(x+h) = \frac{1}{2} \frac{9\cosh[3\beta(x+h)/2] + \cosh[\beta(x+h)/2]\exp(-2\beta D)}{2\cosh[3\beta(x+h)/2] + 2\cosh[\beta(x+h)/2]\exp(-2\beta D)}, \tag{5b}$$

$$F_3(x+h) = \frac{1}{2} \frac{27\sinh[3\beta(x+h)/2] + \sinh[\beta(x+h)/2]\exp(-2\beta D)}{4\cosh[3\beta(x+h)/2] + 4\cosh[\beta(x+h)/2]\exp(-2\beta D)}. \tag{5c}$$

Here, $\beta = 1/k_B T$, T is the absolute temperature, $k_B$ is the Boltzmann constant and $k_B = 1.0$ throughout the paper. By using the definitions of the magnetizations in Eqs. (2a)-(2c), the total magnetization $M_T$ of each site can be defined from Fig. 1(a) as

$$M_T = 1/7\left(6M_S + M_C\right), \tag{6}$$

We should also mention that we did not examine the thermal behaviors of $q_S$, $q_C$, and $r_S$ and $r_C$ due to the reason that our Hamiltonian did not contain the biquadratic exchange interaction parameter as seen in Eq. (1). However, we need Eqs. (3) and (4) to determine the behaviors of $M_S$ and $M_C$.

The internal energy of per site of the system can be calculated as

$$\frac{U}{N} = -\frac{1}{2}\left(\langle U_C \rangle + \langle U_S \rangle\right) - D\left(\langle q_C \rangle + \langle q_S \rangle\right) - h\left(\langle m_C \rangle + \langle m_S \rangle\right), \tag{7}$$

where,

$$U_C = \frac{\partial}{\partial \nabla}\left[c_0 + c_1\langle S_m \rangle + c_2\langle S_m \rangle^2 + c_3\langle S_m \rangle^3\right]^2 \left[b_0 + b_1\langle S_i \rangle + b_2\langle S_i \rangle^2 + b_3\langle S_i \rangle^3\right]^6 F_1(x+h)\bigg|_{x=0}, \tag{8a}$$

$$U_S = \frac{\partial}{\partial \nabla}\left[a_0 + a_1\langle S_i \rangle + a_2\langle S_i \rangle^2 + a_3\langle S_i \rangle^3\right]^4 \left[b_0 + b_1\langle S_m \rangle + b_2\langle S_m \rangle^2 + b_3\langle S_m \rangle^3\right] F_1(x+h)\bigg|_{x=0}. \tag{8b}$$

The specific heat of the system can be obtained from the relation

$$C = \frac{\partial}{\partial \nabla}\left(\frac{\partial U}{\partial T}\right). \tag{9}$$

The Helmholtz free energy of the system can be defined as:

$$F = U - TS \tag{10}$$

in which, according to the third law of thermodynamics, it can be written in the form

$$F = U - T \int_0^T \frac{C}{T'} dT'. \qquad (11)$$

The second term at the right–hand side of Eq. (11) (the integral which appears) is the entropy of the system according to the second law of the thermodynamics.

### 3. Numerical results and discussions

In this section our attention is focused on the study of the magnetic properties, the phase diagrams and hysteresis behavior of a spin-3/2 HIN system with core-shell structure.

*3.1. Magnetic properties*

*3.1.1. Magnetizations*

In Fig. 2, we investigated the thermal behavior of the total ($M_T$), core ($M_C$) and shell ($M_S$) magnetizations both ferromagnetic (r > 0.0) and antiferromagnetic (r < 0.0) cases. This study leads us to characterize of transitions as well as to obtain the transition points. For ferromagnetic and antiferromagnetic cases, a few explanatory examples are plotted in Fig. 2(a) and (b) to illustrate the calculation of the critical points as well as variation of magnetizations. Fig. 2(a) is obtained for r = 1.0, D = 1.0, $\Delta_S$ = 0.0. In this figure, the magnetizations $M_T$, $M_S$, $M_C$ decrease continuously with the increasing of the values of temperature below the critical temperature and they become zero at $T_C$ = 7.85; hence, a second-order phase transition occurs. The transition is from the ferromagnetic-3/2 ($F_{3/2}$) phase to the P phase. Fig. 2(b) is obtained for r = -1.0, D = 1.0, $\Delta_S$ = 0.0. In this figure, the magnetizations $M_T$, $M_S$ decrease and $M_C$ increases continuously with the increasing of the values of temperature below the critical temperature and they become zero at $T_C$ = 7.85; hence, a second-order phase transition occurs. The transition is from the antiferromagnetic-3/2 ($AF_{3/2}$) phase to the P phase.

*3.1.2. Internal energy, specific heat, entropy and Helmholtz free energy*

The Fig. 3(a)-(d) show the thermal behaviors of the internal energy, specific heat, entropy and Helmoltz free energy of the spin-3/2 HIN system, respectively; Fig. 3 obtained for r = 1.0, D = 1.0, $\Delta_S$ = 0.0 values. Fig. 3(a) shows the behavior of the internal energies. It express a discontinuity of the curvature at the critical temperature $T_C$ = 7.85. The specific heat curves of the system exhibit second-order at $T_C$ = 7.85 and rapidly decrease with the increasing temperature, as seen in Fig. 3(b). Fig. 3(c) illustrates the entropy of the spin-3/2 HIN system. As known, entropy is not important at low temperatures and ground state energy correspond to the free energy of the system. But, along with the temperature growing, in order to minimize its free energy the system wants to maximize its entropy. In this way, entropy becomes important. Finally, Helmholtz free energy of the spin-3/2 HIN system is presented in Fig. 3(d). It is well known from Eq. (10) that the free energy equals the internal energy at zero temperature. Free energies, namely total, core and shell free energies, show a smooth decrease with the increasing temperature.

*3.2. Phase Diagrams*

The phase diagrams of the spin-3/2 HIN system are presented in (D, T) plane both ferromagnetic and antiferromagnetic cases in Fig. 4. In these figures, the solid and dashed lines represent the second- and first-order phase transitions, respectively. The system also illustrates a special point, namely double critical end point (B). Fig. 4(a) is calculated for ferromagnetic case (r > 0.0) and r = 1.0, $\Delta_S$ = 0.0. The system shows ferromagnetic-3/2 ($F_{3/2}$), ferromagnetic-1/2 ($F_{1/2}$) and paramagnetic (P) fundamental phases as well as the B special point. The phase boundary is a second-order phase line except for the boundary between the $F_{3/2}$ and $F_{1/2}$ phases which is a first-order line. First-order phase transition line occurs for low D values. Fig. 4(b) is similar to Fig. 4(a) but differs from it: (i) the phase diagram is calculated for (r < 0.0) and r = -1.0, $\Delta_S$ = 0.0. (ii) The $AF_{3/2}$ and $AF_{1/2}$ fundamental phases appear instead of the $F_{3/2}$ and $F_{1/2}$ fundamental phases.

### 3.3. Hysteresis behaviors

Our investigations in this subsection are to examine the effects of the temperature and crystal field on the hysteresis behaviors of a spin-3/2 HIN system.

#### 3.3.1. The effects of temperature on the hysteresis behaviors

We illustrate the dependence of hysteresis loops of a spin-3/2 HIN system on the temperature (T = 2.0, 3.0, 5.0 and 6.3) for r = 0.5, $\Delta_S$ = 0.0 and D = 0.5 in Fig. 5. From this figure can be seen that the magnetization curves are symmetric for both positive and negative values of the external magnetic field. Moreover, for a temperature above the critical temperature $T_C$ = 6.15, we can see that the hysteresis loops disappear. Similar hysteresis loop behaviors have been observed in random field Ising model [41], and the EFT [14, 17, 42, 43]. Furthermore, with the increasing temperature, the hysteresis loops decrease and these results are consistent with some theoretical [44-46] and experimental results [47-52].

#### 3.3.2. The effects of crystal field on the hysteresis behaviors

In order to show the influence of the crystal field on hysteresis, Fig. 6 is plotted for r = 1.0, $\Delta_S$ = 0.0 and T = 2.0 fixed values, and D = 0.0, -1.0, -2.0 and -3.0 crystal field values. We can see that with the decrease of the crystal field, the hysteresis loop area narrows and the system illustrates only one hysteresis loop. These facts are clearly seen in Fig. 6 and the results are consistent with some theoretical [44-46].

#### 3.3.3. The effect of core/shell interfacial coupling on the hysteresis

We present the core/shell interfacial coupling dependence of the hysteresis loops of a spin-3/2 HIN system at T = 2.0, D = 0.0 for $\Delta_S$ = 0.0, and r = 0.05, 0.2, 0.3, 1.2, and 1.8 in Fig. 7. We can see that when the core/shell interfacial coupling constant is large, the system shows always one loop. One can see that with the core/shell interfacial coupling increases, the hysteresis behavior changes loop and the hysteresis loop area is increasing. This fact is clearly see from the Figs. 7 (a)-(e).

## 4. Summary and conclusion

By utilizing the effective-field theory with correlations, we studied magnetic properties such a magnetizations, internal energy, specific heat, entropy, Helmholtz free energy, phase diagrams of a spin-3/2 Blume-Capel model on HIN with core-shell structure. The effects of the Hamiltonian parameters on hysteresis behaviors are investigated, in detail. Finally, the obtained results are compared with some experimental and theoretical results and a qualitatively good agreement is found. Finally, we hope that our detailed theoretical investigations may stimulate further studies on the magnetic properties of nanoparticles systems, and it will also motivate researchers to investigate the behaviors in real nanostructured materials.

**Figure Captions**

**Fig. 1.** (Color online) Schematic representation of hexagonal Ising nanowire. The blue and red spheres indicate magnetic atoms at the surface shell and core, respectively.

**Fig. 2.** (Color online) Thermal variations of the magnetizations with the various values of r, D and $\Delta_S$.
   **(a)** r = 1.0, D = 1.0 and $\Delta_S$ = 0.0,
   **(b)** r = -1.0, D = 1.0 and $\Delta_S$ = 0.0.

**Fig. 3.** (Color online) For r = 1.0, D = 1.0, $\Delta_S$ = 0.0 values, Fig. 3(a)-(d) show the thermal behaviors of the internal energy, specific heat, entropy and Helmoltz free energy of the spin-3/2 HIN system, respectively

**Fig. 4.** The phase diagrams in (T-D) plane of the spin-3/2 HIN system. Dashed and solid lines represent the first- and second-order phase transitions, respectively.

   **(a)** r = 1.0 and $\Delta_S$ = 0.0.
   **(b)** r = -1.0 and $\Delta_S$ = 0.0.

**Fig. 5.** (Color online) Hysteresis behaviors of the spin-3/2 HIN system for r = 0.5, $\Delta_S$ = 0.0, D = 0.5 and for various values of temperatures.

   **(a)** T = 2.0; **(b)** T = 3.0; **(c)** T = 5.0; **(d)** T = 6.3.

**Fig. 6.** (Color online) Same as Fig. (5), but for r = 1.0, $\Delta_S$ = 0.0, T = 2.0, and

   **(a)** D = 0.0; **(b)** D = -1.0; **(c)** D = -2.0; **(d)** D = -3.0.

**Fig. 7.** Same as Fig. (5), but for T = 2.0, D = 0.0 for $\Delta_S$ = 0.0, and

   **(a)** r = 0.05; **(b)** r = 0.2; **(c)** r = 0.3; **(d)** r = 1.2; **(e)** r = 1.8.

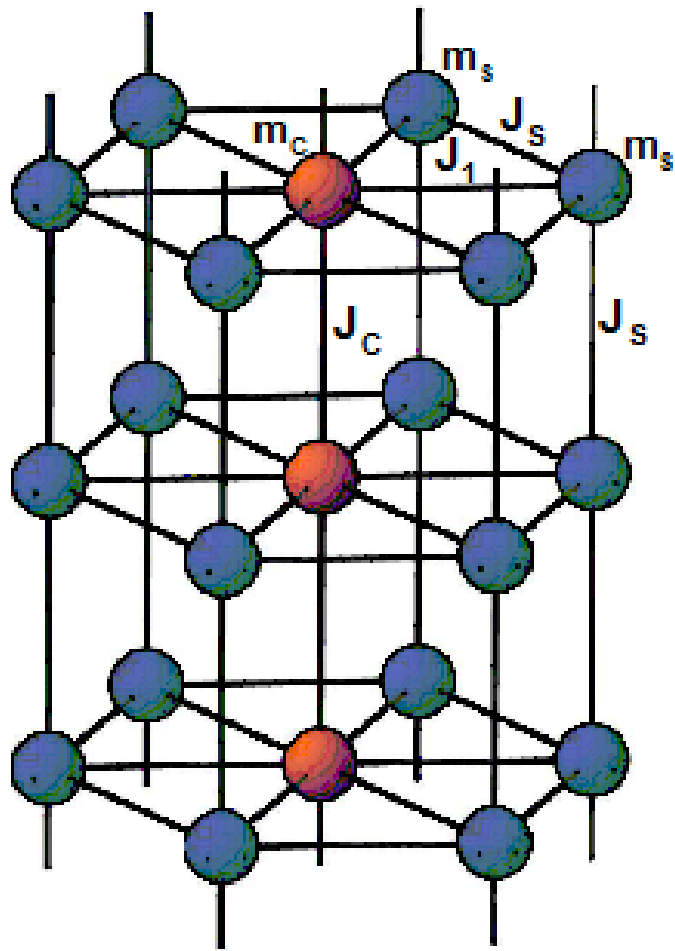

**Fig. 1**

**Figure 2**

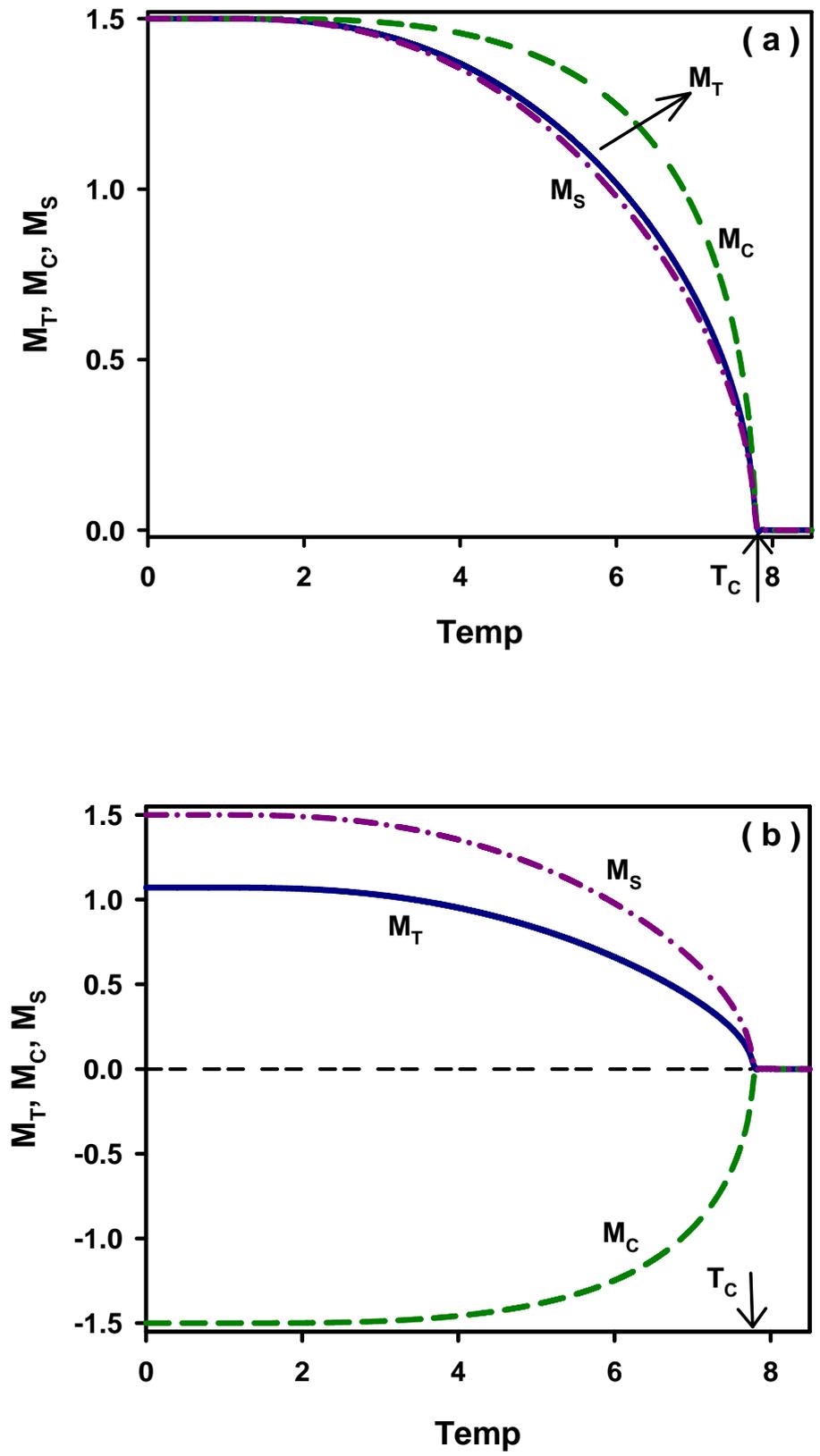

**Fig. 2**

Figure 3

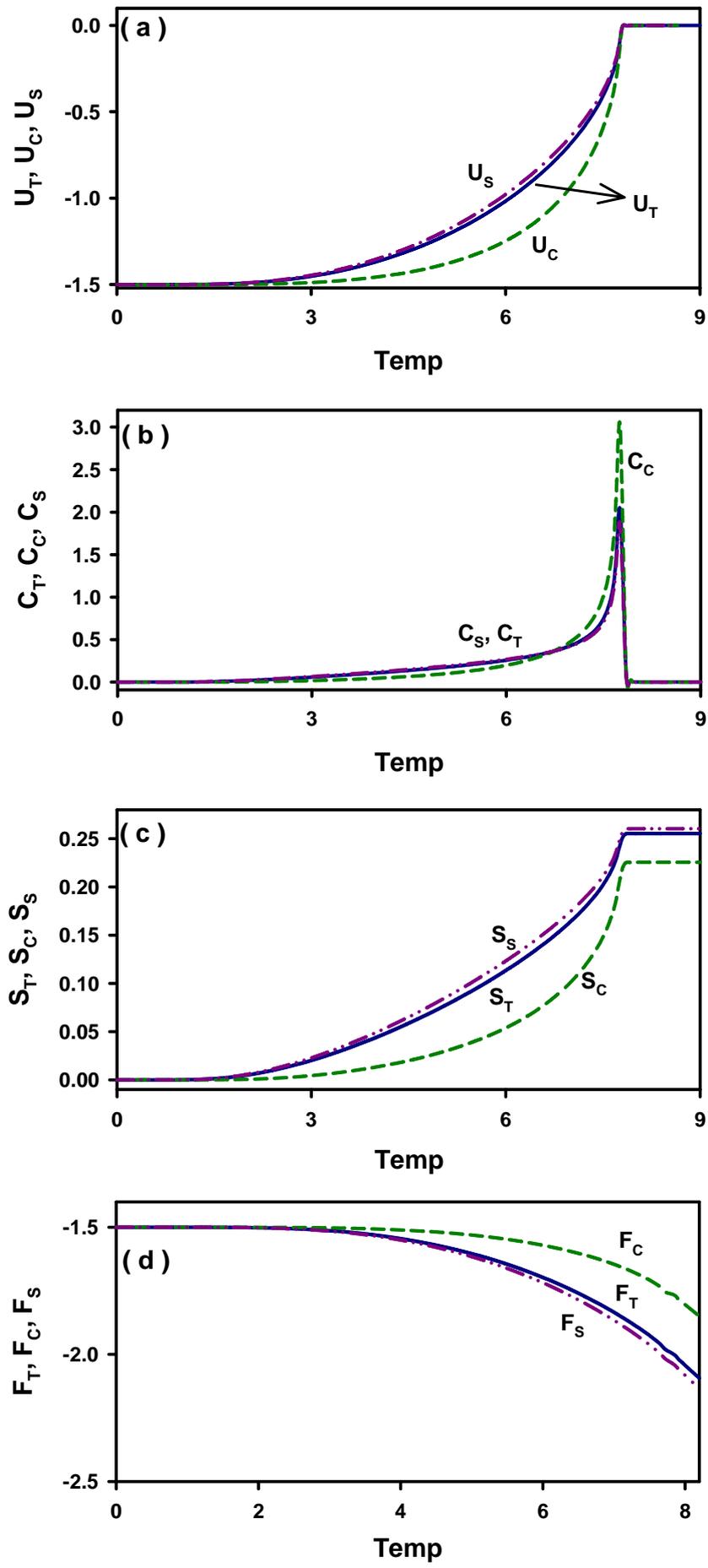

Fig. 3

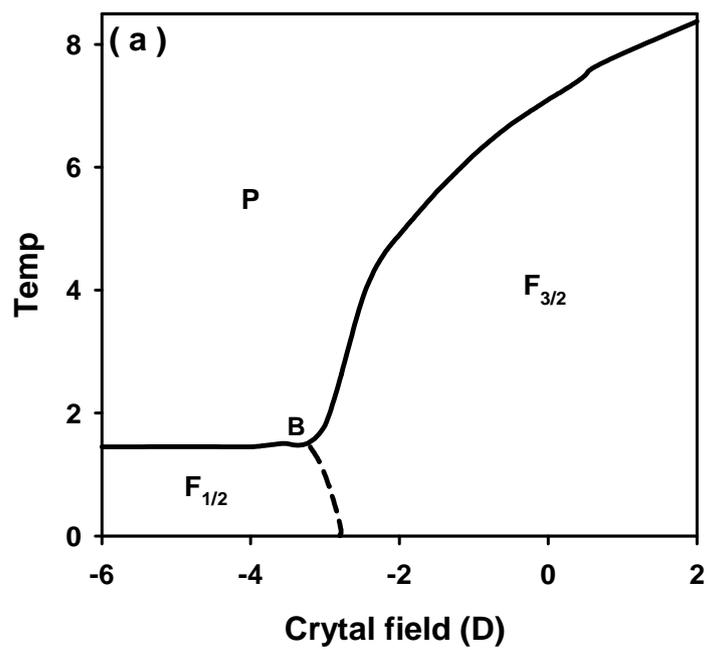

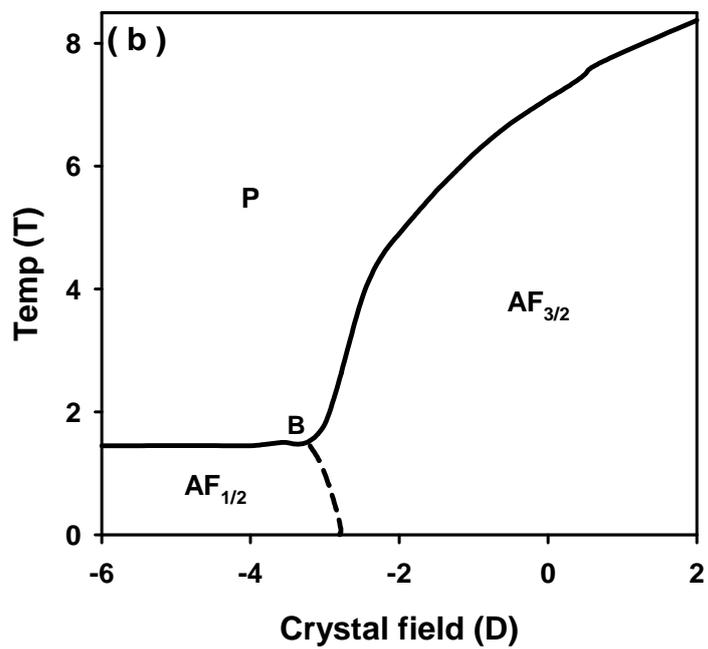

**Fig. 4**

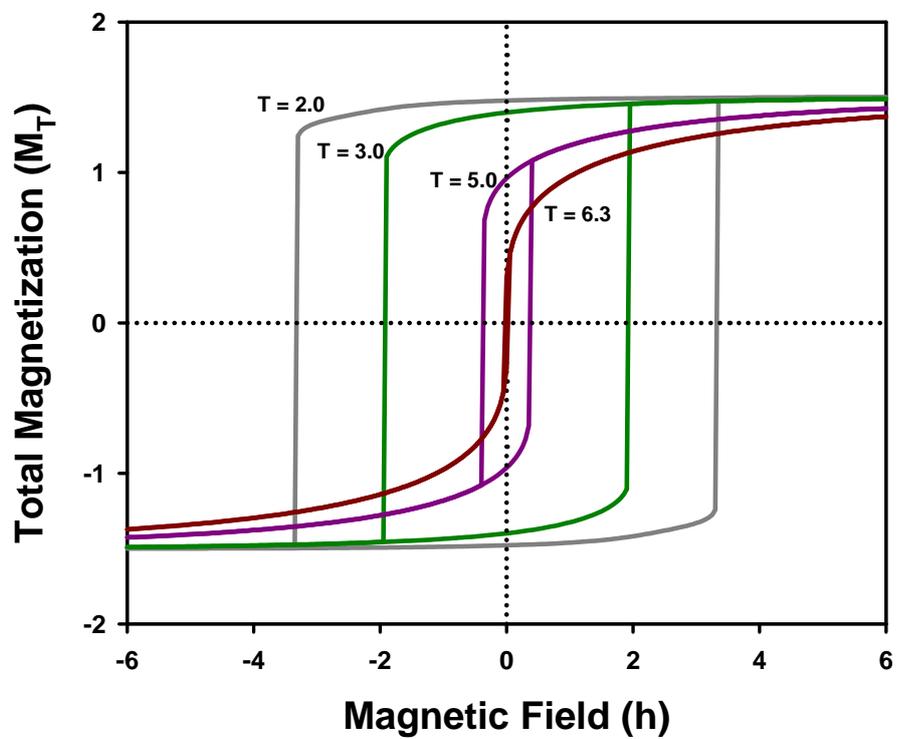

**Fig. 5**

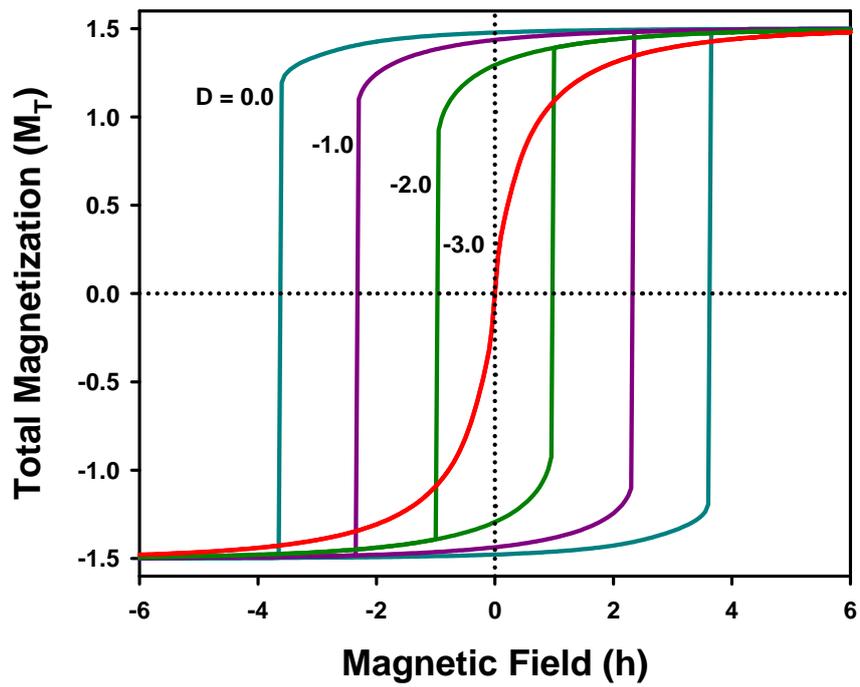

**Fig. 6**

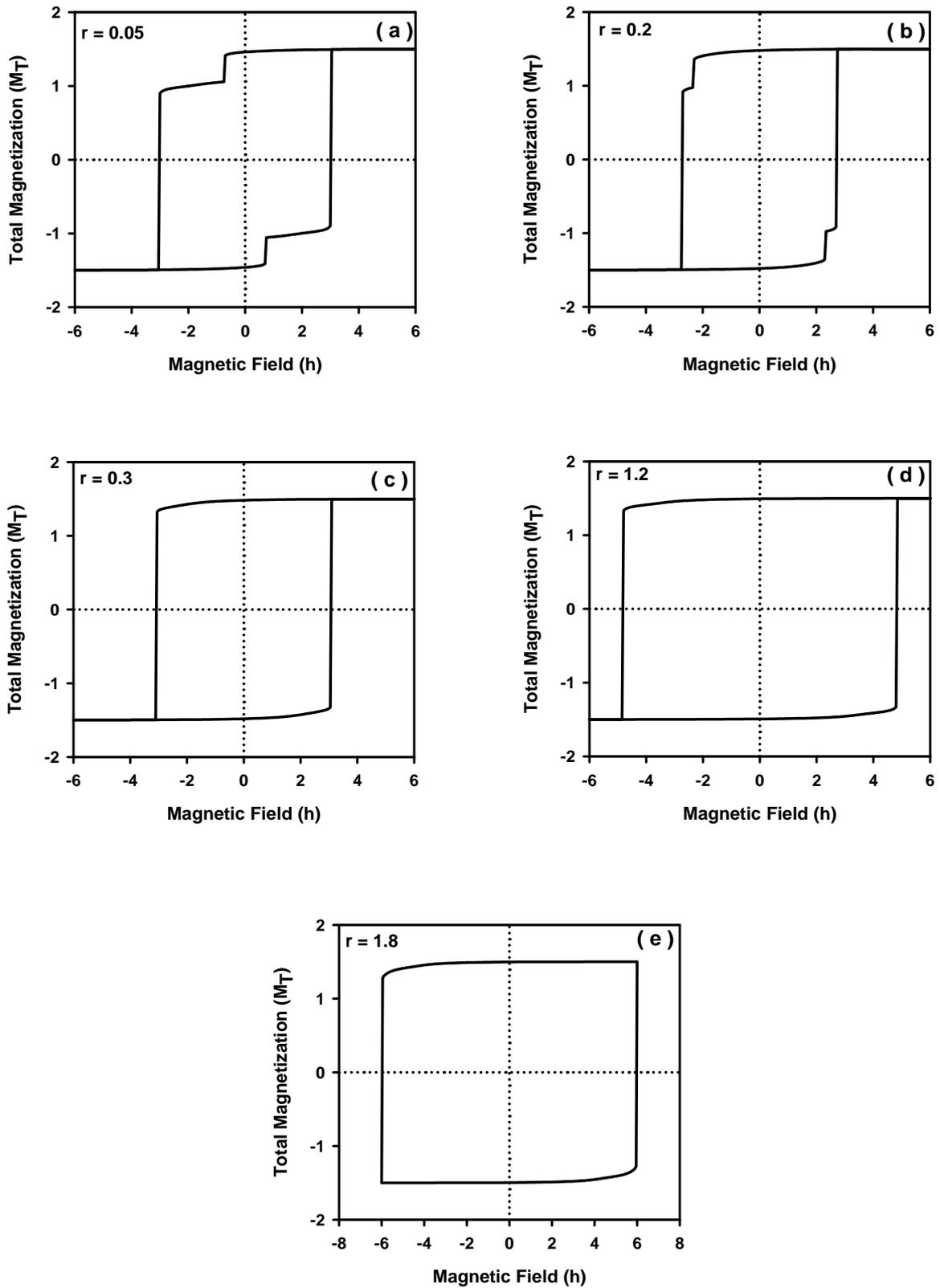

**Fig. 7**

# Research Highlights

- Magnetic properties of a spin-3/2 hexagonal Ising nanowire are studied.
- We examined the effects of the Hamiltonian parameters on the magnetic properties.
- The phase diagrams are obtained in (D-T) plane.
- Hysteresis behaviors are investigated, in detail.